\documentstyle[aps,prl,preprint,floats,epsfig]{revtex}

\def\ifmath#1{\relax\ifmmode#1\else$#1$\fi}
\def\taup	{\ifmath{\tau^{+}}}
\def\taum	{\ifmath{\tau^{-}}}
\def\nutau      {\ifmath{\nu_{\tau}}}
\def\pip	{\ifmath{\pi^{+}}}
\def\pim	{\ifmath{\pi^{-}}}
\def\cbcp   {\ifmath{\cos{\beta}\cos{\psi}}}
\def\thcp   {\ifmath{\theta_{cp}}}
\def\tkspn  {\ifmath{\taum \rightarrow \KS \pim \nutau}}
\def\er #1 #2 { $#1 \pm #2$ }
\def\KS    {\ifmath{K^0_{\scriptscriptstyle S}}} 
\def\KL    {\ifmath{K^0_{\scriptscriptstyle L}}} 
\def\piz   {\ifmath{\pi^0}}

\begin{document}

\preprint{\tighten\vbox{\hbox{\hfil CLNS 98/1557}
                        \hbox{\hfil CLEO 98-7}
}}

\title{First Search for $CP$ violation in Tau Lepton Decay}

\author{CLEO Collaboration}
\date{\today}

\maketitle
\tighten

\begin{abstract} 
We have performed the first search for $CP$ violation in tau lepton decay.
$CP$ violation in lepton decay does not occur in the minimal standard
model but can occur in extensions such as the multi-Higgs doublet model.
It appears as a characteristic difference between the 
$\taum$ and $\taup$ decay angular distributions for the semi-leptonic
decay modes such as $\taum \rightarrow K^{0} \pim \nu$. We define an observable asymmetry to exploit this and find no evidence for any $CP$ violation. 

\end{abstract}
\newpage

{
\renewcommand{\thefootnote}{\fnsymbol{footnote}}

\begin{center}
S.~Anderson,$^{1}$ Y.~Kubota,$^{1}$ S.~J.~Lee,$^{1}$
J.~J.~O'Neill,$^{1}$ R.~Poling,$^{1}$ T.~Riehle,$^{1}$
A.~Smith,$^{1}$
M.~S.~Alam,$^{2}$ S.~B.~Athar,$^{2}$ Z.~Ling,$^{2}$
A.~H.~Mahmood,$^{2}$
% ...
\end{center}
 
\small
\begin{center}
$^{1}${University of Minnesota, Minneapolis, Minnesota 55455}\\
$^{2}${State University of New York at Albany, Albany, New York 12222}\\
% ...
\end{center}

\setcounter{footnote}{0}
}
\newpage


To date $CP$ violation has only been observed in the kaon system~\cite{CCFT}
and its origin remains unknown. In the minimal standard model (MSM) 
$CP$ violation is restricted to the quark sector and cannot occur in lepton
decay~\cite{CKM}. It can, however, occur in extensions to the MSM such
as the three Higgs doublet model~\cite{SW}. It appears that there 
is insufficient $CP$ violation in the MSM to generate the apparent
matter-antimatter asymmetry of the universe~\cite{KT}. Searches for 
additional $CP$ violation beyond the MSM may help reconcile this problem.

$CP$ violation appears as a phase  $\thcp$ in the gauge boson-fermion coupling
constant, $CP$:$\thcp \rightarrow -\thcp$. The physical effects of such a 
phase are only manifest in the interference of two amplitudes with both 
relative $CP$-odd phase $\thcp$ and relative $CP$-even phase $\delta$ (the interference term is proportional to  $\cos{(\delta - \thcp)}$). In tau lepton decay the
two amplitudes could come from the MSM vector boson exchange ($W$) and 
the extended standard model scalar (Higgs) exchange. The $CP$-odd phase comes 
from the imaginary part of the complex scalar coupling constant. The $CP$-even
 phase difference is provided by the final state interaction (strong) phase
that is different for $s$-wave scalar exchange  and $p$-wave vector exchange 
and only arises in semi-leptonic decay modes with at least
two final state hadrons ($\taum \rightarrow h_{1}h_{2}\nutau$). The final state interaction is described by the $s$-wave
and $p$-wave form factors, $F_{s}=|F_{s}|e^{i\delta_{s}}$ and
 $F_{p}=|F_{p}|e^{i\delta_{p}}$ respectively so that the strong phase
difference is $\delta_{strong}=\delta_{p}-\delta_{s}$. The $CP$-violating
$s-p$ wave interference term is then proportional to  
$|F_{p}||F_{s}|g\cos{(\delta_{strong} - \thcp)}\cbcp $, where $\beta$ and $\psi$
are physical decay angles measured in the hadronic
rest frame ($\vec{p}_{h_{1}}+\vec{p}_{h_{2}}=0$)~\cite{KM}. The direction of the
laboratory frame as viewed from the hadronic rest frame is $\vec{p}_{lab}$
and $\beta$ is the angle between the direction of $h_{1}$ or $h_{2}$  and $\vec{p}_{lab}$. $\psi$
is the angle between the tau flight direction and $\vec{p}_{lab}$. The
ratio of scalar to vector coupling strength is $g$ (i.e $g$ is in units of $G_{F}/2\sqrt{2}$).
Since the sign of $\thcp$ changes for the $CP$-conjugate $\taum$ and
$\taup$, we define an experimentally measurable asymmetry 
$A_{observed}(\cbcp)$ in terms of the number of events from $\tau^{\pm}$
decay, $N^{\pm}(\cbcp)$, in a particular interval of $\cbcp$:
\[
A_{observed}(\cbcp)=\frac{N^{+}(\cbcp)-N^{-}(\cbcp)}
            {N^{+}(\cbcp)+N^{-}(\cbcp)}
  \propto   |F_{p}||F_{s}|g\sin{\delta_{strong}}\sin{\thcp}\cbcp
\] 
The asymmetry is linear in $\cbcp$ and we do not expect an overall rate
asymmetry~\cite{EXP} since 
\[
\int_{-1}^{+1}dx A_{observed}(x) =0  ; x=\cbcp
\]

We select a $\taum \rightarrow \KS h^{-} \nutau$, $\KS \rightarrow \pip \pim$ event sample 
since a mass dependent Higgs-like coupling would give the largest asymmetry
in this mode and with three charged tracks in the final state the decay 
angles are well measured. Here $h^{-}$ is a charged pion or kaon.

   The data used in this analysis have been collected from $e^{+}e^{-}$
collisions at a center of mass energy ($\sqrt{s}$) of 10.6 GeV with the 
CLEO II detector at the Cornell Electron Storage Ring (CESR). The total
integrated luminosity of the data sample is 4.8 fb$^{-1}$, corresponding
to the production of $4.4 \times 10^{6}$ $\taup\taum$ events. The
CLEO II detector has been described elsewhere~\cite{CD}.

   We select events with a total of 4 charged tracks and zero net charge. Each track
must have  momentum transverse
to the beam axis $p_{T} >  0.025E_{beam}$($E_{beam}=\sqrt{s}/2$) and 
$|\cos{\theta}|< 0.90$ where $\theta$ is the polar angle with respect to 
the beam direction. The
event is divided into two hemispheres by requiring one of the charged
tracks to be isolated by at least 90$^{o}$ from the other three
(1 vs 3 topology). The isolated track is then required to have 
momentum greater than 0.05$E_{beam}$ and $|\cos{\theta}|< 0.80$
to ensure efficient triggering and reduce backgrounds from two photon
processes and beam gas interactions. To further reduce the two
photon backgrounds and also continuum quark-antiquark production ($q\overline{q}$) we 
require that the net
missing momentum of the event be greater than $0.03E_{beam}$ in the
transverse plane and not point to within $18^{o}$ of the beam axis.
We also require the total visible energy in the event to be between
$0.7E_{beam}$ and $1.7E_{beam}$.
 
 Events are permitted to contain a pair of {\it unmatched} energy
clusters in the calorimeter (i.e, those not matched with a charged 
particle projection) in the 1-prong hemisphere with energy greater
than 100 MeV consistent with $\piz$ decay. After $\piz$ 
reconstruction we reject events with remaining unmatched showers 
of greater than 350 MeV. We further reject events with showers of energy
above 100 MeV in the 3 prong hemisphere or 300 MeV in the 1-prong
hemisphere provided such showers are well isolated from the nearest 
track projection (by at least  30 cm) and have photon-like 
lateral profiles. These vetoes suppress backgrounds from $q\overline{q}$
events and tau feed-across (i.e., tau decay modes containing unreconstructed
$\piz$'s or $\KL$'s).

 The $\KS$ is identified by requiring two of the tracks in the 3-prong
hemisphere to be consistent with the decay  $\KS \rightarrow \pip\pim$. We determine the $\KS$ decay point in the x-y plane (transverse to the beam direction) by the
intersection of the two tracks projected onto this plane. This point must lie at
least 5 mm from the mean $e^{+}e^{-}$ interaction point (IP). We require that
the distance between the two tracks  in $z$ (beam direction) at the decay point be less than 
12 mm to ensure that the tracks form a good vertex in three dimensions. The
distance of closest approach to the IP of the line defined by the x-y
projection of the $\KS$ momentum vector must be less than 2 mm. The
invariant mass of the pair of tracks, assumed to be pions, must be within
20 MeV of the known $\KS$ mass. We define a sideband region 30-100 MeV
above and below the $\KS$ mass to use as a control sample.

\makebox[368bp]{\epsfysize=3.5in \epsfxsize=5.0in \epsffile{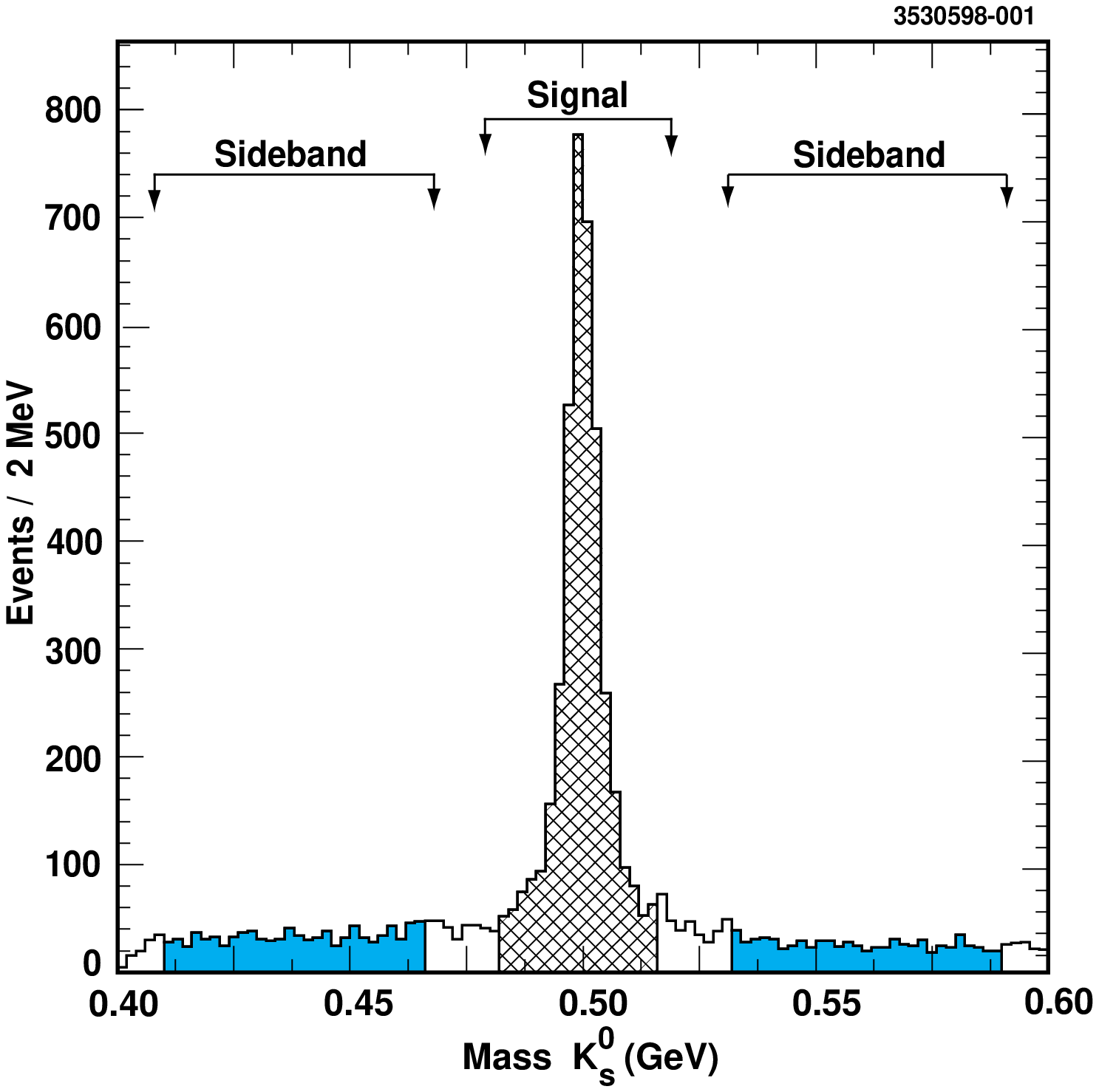}}
\begin{figure}[h]
\vspace{0.1in}
       \caption{Invariant mass distribution for $\KS \rightarrow \pim \pip$ in 
final data sample.}
\label{fig1}
\end{figure}

Figure~\ref{fig1} shows the invariant
mass distribution after all selection criteria. Using this sample 
we measure the asymmetry for both signal and sideband in two intervals of 
$\cbcp$, $A_{observed}
(\cbcp < 0)$ and $A_{observed}(\cbcp > 0)$, given in
Table~\ref{tab1}. 
Both signal and sideband exhibit similar non-zero asymmetries
but with low statistical significance. The measured asymmetries are
insensitive to small variations in the selection criteria. In addition to
CP-violation, a non-zero asymmetry can arise from either a statistical
fluctuation or a difference in detection efficiency for positive and
negatively charged particles. A Monte Carlo simulation is used to estimate the
expected $CP$ violation in terms of the extended standard model scalar
coupling parameters and we use the sideband sample to empirically estimate
the asymmetry due to detector effects.

\begin{table}[htb]
\begin{center}
\begin{tabular}{|l|c|c|}\hline
                & $A_{observed}(\cos{\beta}\cos{\psi}<0)$ & $A_{observed}(\cos{\beta}\cos{\psi}>0)$ \\ \hline
Signal          &  \er 0.058 0.023   & \er 0.024 0.021    \\ \hline
Sideband        &  \er 0.049 0.030   & \er 0.034 0.033    \\ \hline
\end{tabular}
\end{center}
\caption{Observed asymmetries in signal and sideband regions}
\label{tab1}
\end{table}

To estimate the expected $CP$-violating  asymmetry 
for a pure $\taum \rightarrow \KS \pim \nutau $ sample we use the KORALB 
Monte Carlo~\cite{JKW} to generate $\tau$-pairs.
It has been modified to include a scalar Higgs
coupling in addition to the standard model $W$ boson coupling, for
the signal $\KS\pim$. 
We set $F_{s}=1$ (i.e non-resonant decay) and $F_{p}$ to be a Breit-Wigner
for the $K^{*}(892)$ resonance so that $F_{p} >> F_{s}$ and the average
strong phase difference $<\delta_{strong}>=\pi/2$. The  GEANT code ~\cite{GE} is used to simulate
detector response and assumes equal detection efficiencies for positive
and negatively charged particles. We estimate $A^{\KS\pim}_{expected}(\cbcp < 0)=-0.033g\sin{\thcp}$ and
$A^{\KS \pim}_{expected}(\cbcp > 0)=+0.033g\sin{\thcp}$ for a pure $\tkspn$
signal. To compare this estimated asymmetry to the observed asymmetry 
we must take into account the effect of backgrounds since the signal region 
is not pure $\KS \pim$ and also estimate the asymmetry expected from charge dependent
detection inefficiencies alone. Table~\ref{tab2} gives the estimated signal and sideband compositions
by mode where the Lund Monte Carlo~\cite{LU} has been used to generate the 
$q\overline{q}$ events.

\begin{table}[htb]
\begin{center}
\begin{tabular}{|l|l|l|l|l| }\hline
Tau       Mode             & $\alpha_{mode}$ & $f_{mode}^{signal}$ & $f_{mode}^{sideband}$ & $(f_{mode}^{signal}-f_{mode}^{sideband}).\alpha_{mode}$   \\ \hline
$\KS(\pip\pim)\pim\nutau$  &    1     & \er 0.525 0.057  &   \er 0.043 0.005     & \er 0.4820   0.0570     \\
$\KS K^{-}\nutau$         &    1/20  & \er 0.124 0.036  &   \er 0.009 0.003     & \er 0.0060  0.0020   \\    
$a_{1}^{-} \nutau$                &    1/80  & \er 0.106 0.003  &   \er 0.620 0.013     & \er -0.0064 0.0002  \\
$\KS \pim\piz\nutau$     &    1/4   & \er 0.066 0.016  &   \er 0.006 0.002     & \er 0.0150  0.0040   \\ 
$\KS\KL\pim\nutau$&    1/80  & \er 0.055 0.018  & \er 0.003 0.001     & \er 0.0007    0 .0002    \\
$\KS K^{-}\piz\nutau$        &    1/20  & \er 0.030 0.008  & \er 0.003 0.001     & \er 0.0014    0.0004    \\
$\pip\pim\pim\piz\nutau$      &    1/20  & \er  0.028 0.002 & \er 0.167 0.007  & \er -0.0070    0.0004    \\      
$K^{-}\pip\pim\nutau$      &    1/4   & \er 0.008 0.003 &   \er 0.043  0.007    & \er -0.0090 0.0020  \\ 
others                     &     0    & \er 0.012 0.002  &   \er 0.071  0.017  &        0          \\
$q\overline{q}$            &     0    & \er 0.044 0.003 &   \er 0.037  0.003  &        0          \\ \hline
Total                      &     -    & \er 1.00   0.07  &   \er 1.00    0.00  & \er 0.48 0.06   \\ \hline
 \end{tabular} 
\end{center}
\caption{Signal and sideband mode composition. $f^{signal,sideband}_{mode}$ is the fraction of the total signal or sideband sample for a particular mode. 
$\alpha_{mode}$ is the approximate magnitude of asymmetry expected relative
to the $\tkspn$ mode. The last column gives the dilution factor expected
in the asymmetry when the measured asymmetry in the sideband control sample
is subtracted from the measured asymmetry in the signal sample.}
\label{tab2}
\end{table}

The backgrounds arise from our inability to distinguish kaons and pions in the
desired momentum range, lack of $\KL$ identification, particles that fall
outside the fiducial region of the detector, and charged track mismeasurement.
We note that the signal and sidebands are 
composed of different modes and it is unlikely that both samples would exhibit a similar $CP$-asymmetry as the strong phases, and possibly the coupling
strengths are different for each mode. Also the samples exhibit an overall
rate asymmetry not expected from $CP$-violating interference effects~\cite{EXP}.
However the effect of charge dependent detection
inefficiencies would be expected to be similar as both samples satisfy the
same kinematic selection criteria.

 Studies of pions from an independent $\KS \rightarrow \pip \pim$ 
sample indicate that at low momentum  the reconstruction efficiency for $\pip$ is slightly greater than $\pim$ and also the reconstruction of a $\KS$ in close proximity to a $\pip$ is slightly more efficient than for a $\pim$. 
The hadronic interaction of $\pip$ with the
CsI crystals produces more fake electromagnetic clusters than from a $\pim$ which may then be used as veto clusters. These effects are more pronounced
at lower momentum ($<$ 1 GeV) and thus
for $\cbcp < 0.0$ since the pion from $\tkspn$ tends to be of lower
momentum in this region. The sidebands may be used as a control sample to estimate
these combined effects in our signal region in a simple empirical way providing 
we assume that any $CP$-violating effects are suppressed in the sideband modes.

The samples consist 
of a sum of modes, each a fraction $f^{signal,sideband}_{mode}$ of the total signal or sideband sample, with a 
possible $CP$-violating asymmetry suppressed by a factor $\alpha_{mode}$ relative to the 
$\tkspn$ signal mode. 
The suppression factor $\alpha_{mode}$ arises from two effects and is
given for each mode in Table~\ref{tab2}. First 
from the mass dependence of the Higgs coupling and second due to the dilution
of the $p$-wave nature of the standard model final state. For example, the
$\taum \rightarrow \pim \pip \pim \nutau$ mode is dominated
in the standard model decay by an $s$-wave $\taum \rightarrow a_{1}^{-} \nutau \rightarrow \rho^{0} \pim \nutau$ intermediate state which dilutes the $s-p$ wave
interference by a factor of $\approx 4$ in addition to a mass suppression of $m_{u}/m_{s}$ relative to the $\KS \pim$ mode. 

If we assume that in the absence of any true $CP$
violation a charge dependent detector inefficiency would produce an asymmetry $A_{detector}$
common to all modes then the observed asymmetry, assuming the asymmetries
are small (i.e $\ll$ 1), is given by
\[
A^{signal,sideband}_{observed}=
\Sigma_{mode}f^{signal,sideband}_{mode}\alpha_{mode}A^{\KS\pim}_{expected} 
+ A_{detector}
\]
From Table~\ref{tab2} we see that the sideband should have negligible
asymmetry with respect to the signal under the assumption of a mass
dependent coupling and can be used as a control sample to subtract the
charge dependent detector asymmetries common to both signal and sideband.
\[
A^{subtracted}_{observed}=
A^{signal}_{observed}-A^{sideband}_{observed}=0.48A^{\KS \pim}_{expected}
\]
If a true $CP$ violation exists the subtracted quantity should still
exhibit significant but diluted asymmetry while detector effects
should be removed. From Table~\ref{tab1} the measured subtracted 
asymmetry is $ A^{subtracted}_{observed}(\cbcp < 0)
=  0.009 \pm 0.038 $ , $A^{subtracted}_{observed}(\cbcp > 0 ) =  -0.010 \pm 0.039 $ 
which is consistent with no $CP$ violation. 
This can be compared
with a revised Monte Carlo estimate that takes into account
the  dilution factor of 0.48, $A_{expected}(\cbcp < 0)= -0.016g\sin{\thcp}$, 
$A_{expected}(\cbcp > 0)= 0.016g\sin{\thcp}$. 
To cross check  
our assumption of suppressed $CP$ violation in the sidebands
we measure the asymmetry in an independent high-purity
high-statistics  data sample of the dominant sideband
mode, $\taum \rightarrow a_{1}^{-} \nutau$, using the selection criteria of reference~\cite{RB}. We find $A^{a_{1}}_{observed}(\cbcp <0)=-0.0013 \pm 0.0047,
 A^{a_{1}}_{observed}(\cbcp > 0)=-0.0023 \pm 0.0047$ 
giving no evidence for $CP$ violation. The higher track momentum and cluster veto thresholds
combined with the absence of a $\KS$
requirement from this sample removes the contribution to the  asymmetry from
charge dependent detection inefficiencies but a true $CP$-violating
effect should remain. We note that by measuring the $CP$-violating
asymmetry in the dominant sideband mode as zero our results are
approximately valid for a non-mass dependent coupling. However, we cannot fully
relax this assumption due to the difficulty of empirically isolating a 
sample of each background mode in which to measure the asymmetry.

In conclusion we find no evidence for $CP$ violation
in tau decay. We may compare the observed
to expected asymmetries to set a constraint of $g\sin{\thcp}< 1.7$ at the 90 \% confidence
level assuming $F_{s}=1$. At the forthcoming $B$-factory experiments we anticipate substantial 
improvements in sensitivity both from the increased statistical
precision and detector improvements. The addition of $\KL$ detection,
$K^{-}/\pim$ separation and improved precision tracking will significantly
decrease backgrounds.

We gratefully acknowledge the effort of the CESR staff in providing us with
excellent luminosity and running conditions.
This work was supported by 
the National Science Foundation,
the U.S. Department of Energy,
Research Corporation,
the Natural Sciences and Engineering Research Council of Canada, 
the A.P. Sloan Foundation, 
the Swiss National Science Foundation, 
and the Alexander von Humboldt Stiftung.


\begin{thebibliography}{99}
\bibitem{EXP}
In the most general case we do not expect an overall rate asymmetry.
The angular dependence of an $s,p,d$ etc.,-wave amplitude is given by the
Legendre polynomials $P_{m=0,1,2}(x)$ so that the angular dependence
of a $CP$-violating interference term is given by $P_{m}(x)P_{n}(x)$ 
and since 
\[
\int_{-1}^{+1}dx P_{m}(x)P_{n}(x) = \frac{2}{2n+1}\delta_{mn}
\]
we can only have a non-zero rate asymmetry if $m=n$ in which case the strong
phases of the two amplitudes will be equal ($\sin{\delta_{strong}}=0$) so
that the overall rate asymmetry will be zero.
\end{thebibliography}
\end{document}